\documentclass[aps,twocolumn,amsmath,amssymb,superscriptaddress,prb]{revtex4-1}

\usepackage{graphicx}
\usepackage{color}

\usepackage{bm}

\usepackage[normalem]{ulem}

\begin{document}

\title{Observation of the orbital Nernst effect}

\author{Yuto Masuda}
\affiliation{Department of Applied Physics and Physico-Informatics, Keio University, Yokohama 223-8522, Japan}

\author{Takamasa Hirai}
\affiliation{National Institute for Materials Science, Tsukuba 305-0047, Japan}

\author{Daegeun Jo}
\affiliation{Department of Physics and Astronomy, Uppsala University, Uppsala SE-75120, Sweden}
\affiliation{Wallenberg Initiative Materials Science for Sustainability, Uppsala University, Uppsala SE-75120, Sweden}

\author{Naoki Yano}
\affiliation{Department of Applied Physics and Physico-Informatics, Keio University, Yokohama 223-8522, Japan}

\author{Peter M. Oppeneer}
\affiliation{Department of Physics and Astronomy, Uppsala University, Uppsala SE-75120, Sweden}
\affiliation{Wallenberg Initiative Materials Science for Sustainability, Uppsala University, Uppsala SE-75120, Sweden}

\author{Hossein Sepehri-Amin}
\affiliation{National Institute for Materials Science, Tsukuba 305-0047, Japan}

\author{Ken-ichi Uchida}
\affiliation{National Institute for Materials Science, Tsukuba 305-0047, Japan}
\affiliation{Department of Advanced Materials Science, Graduate School of Frontier Sciences, The University of Tokyo, Kashiwa 277-8561, Japan}

\author{Kazuya Ando\footnote{Correspondence and requests for materials should be addressed to ando@appi.keio.ac.jp}}
\affiliation{Department of Applied Physics and Physico-Informatics, Keio University, Yokohama 223-8522, Japan}
\affiliation{Keio Institute of Pure and Applied Sciences, Keio University, Yokohama 223-8522, Japan}
\affiliation{Center for Spintronics Research Network, Keio University, Yokohama 223-8522, Japan}

\maketitle

\bigskip\noindent
\textbf{Abstract}

\textbf{
  The Nernst effect, which converts a temperature gradient into a transverse charge current, is fundamental to thermoelectrics. Its spin analogue, the spin Nernst effect, enables thermal generation of transverse spin currents and is central to spin caloritronics. Recently, the discovery of orbital currents---the orbital counterpart of spin currents---has extended angular-momentum transport beyond spin, leading to the prediction of the orbital Nernst effect, in which a temperature gradient drives a transverse orbital current. However, experimental evidence for this effect has been lacking. Here, we report the observation of the orbital Nernst effect in Ti. Using Ni electrodes on Ti, we detect a thermally induced voltage that depends on the magnetization direction and scales linearly with the temperature gradient. This voltage is strongly suppressed both when Ni is replaced with Ni$_{81}$Fe$_{19}$ and when Ti is replaced with Cr, providing strong evidence that the signal originates from the orbital Nernst effect rather than the anomalous Nernst or spin Nernst effect. These results establish thermally driven orbital transport, opening a pathway toward orbital caloritronics.
}

\bigskip\noindent
\textbf{Main}

\begin{figure*}[tb]
\includegraphics[scale=1]{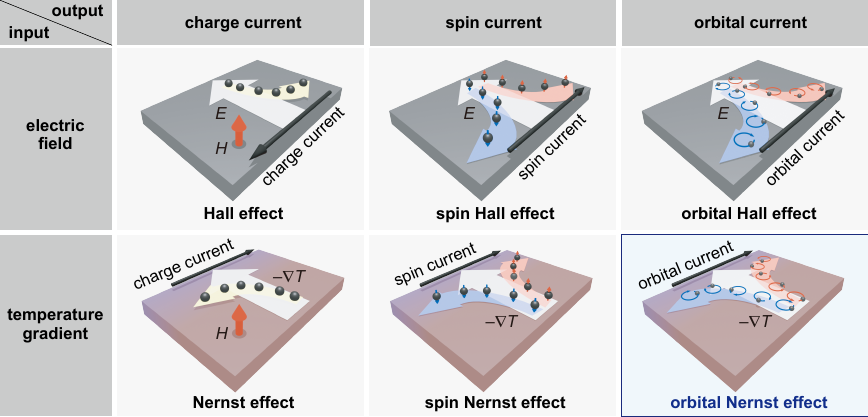}
\caption{
\textbf{Electrical and thermal generation of charge, spin, and orbital currents.}
Top row: Electrical generation via the Hall effect (left), the spin Hall effect (center), and the orbital Hall effect (right), where an applied electric field $E$ induces a transverse charge current $j_\mathrm{C}$, spin current $j_\mathrm{S}$, or orbital current $j_\mathrm{L}$, respectively.
Bottom row: Thermal generation via the Nernst effect (left), the spin Nernst effect (center), and the orbital Nernst effect (right), where a temperature gradient $\nabla T$ drives a transverse charge current $j_\mathrm{C}$, spin current $j_\mathrm{S}$, or orbital current $j_\mathrm{L}$, respectively.
The spin and orbital effects are illustrated for positive spin and orbital Hall and Nernst conductivities.
}
\label{figHallNernst}
\end{figure*}

The ability to generate and manipulate spin currents has driven the rapid development of spintronics~\cite{oxfordspin}. One of the key phenomena is the spin Hall effect, where a charge current generates a transverse spin current in materials with strong spin-orbit coupling~\cite{RevModPhys.87.1213} (see Fig.~\ref{figHallNernst}). By enabling electrical generation of spin currents, the spin Hall effect has played a crucial role in exploring both the physics and applications of spin currents, ranging from the discovery of spin-orbit torques and spin Hall magnetoresistance to the realization of nonvolatile magnetic memories and spin Hall nano-oscillators~\cite{AndoPRL,PhysRevLett.106.036601,PhysRevLett.110.206601,RevModPhys.91.035004,awad2017long}.

Beyond electrical driving, temperature gradients can also generate spin currents~\cite{bauer2012spin,boona2014spin,uchida2021transport}. The thermal analogue of the spin Hall effect is the spin Nernst effect, in which a temperature gradient drives a transverse spin current through the spin Hall mechanism even in the absence of a longitudinal charge current~\cite{meyer2017observation,sheng2017spin,kim2017observation,bose2018direct,park2021geometrical,wimmer2021low,jain2023thermally,li2025large} (see Fig.~\ref{figHallNernst}). The spin Nernst effect therefore provides a purely thermal pathway to generate spin currents and has emerged as a central phenomenon in spin caloritronics, the field focused on heat-spin coupled transport.

Recent advances have revealed that, in addition to spin currents, orbital currents---the flow of orbital angular momentum of electrons---play a fundamental role in angular momentum transport in solids~\cite{go2021orbitronics,KIM2022169974,jo2024spintronics,wang2025orbitronics,2025Andoreview}. The orbital counterpart of the spin Hall effect, the orbital Hall effect, has been demonstrated to generate a transverse orbital current under an applied electric field, even in systems without spin-orbit coupling (see Fig.~\ref{figHallNernst})\cite{PhysRevLett.95.066601,PhysRevB.77.165117,PhysRevLett.102.016601,PhysRevLett.121.086602,PhysRevB.98.214405,PhysRevLett.126.056601,PhysRevMaterials.6.095001}. Orbital analogues of spin phenomena, including orbital torque, orbital pumping, and magnetoresistance effects induced by orbital currents, have also been experimentally observed~\cite{Cr-orbital,Ta-orbital,PhysRevResearch.4.033037,hayashi2022observation,choi2021observation,PhysRevResearch.5.023054,PhysRevLett.125.177201,PhysRevLett.128.067201,PhysRevLett.131.156702,PhysRevB.108.144436,moriya2024nano,BulkSantos2024,gao2024control,hayashi2024observation,hayashi2023orbital,el2023observation,xu2024orbitronics,seifert2023time}, establishing the emerging field of orbitronics~\cite{go2021orbitronics,KIM2022169974,jo2024spintronics,wang2025orbitronics,2025Andoreview}. This progress extends the scope of angular momentum transport beyond spin and has led to the prediction of a thermal analogue of the orbital Hall effect: the orbital Nernst effect, in which a temperature gradient drives a transverse orbital current~\cite{PhysRevMaterials.6.095001} (see Fig.~\ref{figHallNernst}). However, experimental evidence for this effect has so far been missing.

In this work, we report the experimental observation of the orbital Nernst effect in Ti. By analogy with the spin Nernst effect, the orbital Nernst effect can originate from heat-driven orbital Hall mechanisms~\cite{PhysRevMaterials.6.095001}. Because the orbital Hall effect does not rely on spin-orbit coupling, the orbital Nernst effect can be substantial even in light metals with weak spin-orbit coupling. 
Indeed, the orbital Nernst effect arising from the energy derivative of the orbital Hall conductivity is predicted to be significant in Ti despite its weak spin-orbit coupling, whereas the corresponding intrinsic spin Nernst effect is predicted to be negligible~\cite{PhysRevMaterials.6.095001}.
Using ferromagnetic Ni contacts as orbital-to-spin converters and spin-accumulation detectors, we observe a magnetic-field-dependent voltage signal that is consistent with the injection of orbital currents at the Ti/Ni interface. We show that replacing Ti with Cr strongly suppresses the voltage signal, providing evidence that the heat-induced voltage observed in the Ti/Ni device cannot be attributed to thermoelectric effects generated within the Ni detector. We further demonstrate that, in Ti-based devices, the signal is strongly suppressed when the Ni detector is replaced with Ni$_{81}$Fe$_{19}$. This suppression is opposite to the trend expected for a spin Nernst origin but is consistent with an orbital Nernst origin, providing key experimental evidence that the observed signal is dominated by the orbital Nernst effect in Ti. These findings support the thermal generation of transverse orbital currents through the orbital Nernst effect, paving the way for the development of orbital caloritronics.

\bigskip\noindent
\noindent 
\textbf{Orbital Nernst effect in Ti}

\begin{figure}[tb]
\includegraphics[scale=1]{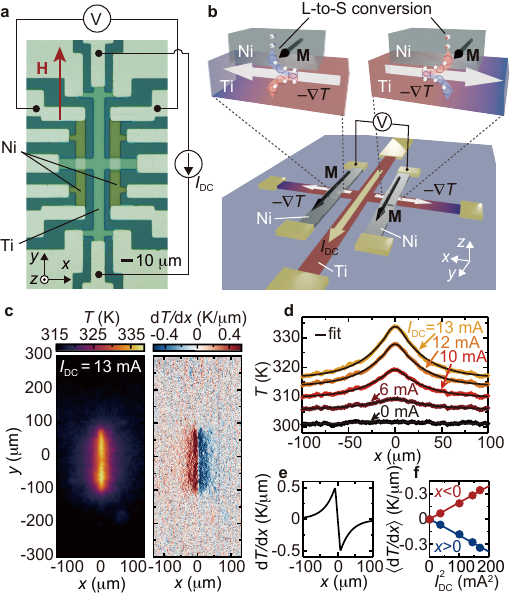}
\caption{
\textbf{Experimental setup and temperature distribution.}
\textbf{a} Optical image of the fabricated device. The central cross-bar structure consists of Ti, while the two lines on the left and right sides are Ni. Bright yellow regions indicate electrodes.
\textbf{b} Schematic of the experimental setup. A DC current $I_\mathrm{DC}$ is applied to the Ti line, generating a temperature gradient $\mathrm{d}T/\mathrm{d}x$ along the $x$ axis. The orbital current generated by the orbital Nernst effect flows along the $z$ axis, with the orbital Nernst conductivity assumed to be negative. A magnetic field $H$ is applied along the $y$ axis. Orbital-to-spin (L-to-S) conversion in the Ni layers is schematically illustrated.
\textbf{c} Spatial distributions of the temperature $T$ and the in-plane temperature gradient $\mathrm{d}T/\mathrm{d}x$ measured by high-resolution infrared thermography in the Ti/Ni device at $I_\mathrm{DC}=13$~mA.
\textbf{d} Temperature profiles $T$ along the $x$ axis obtained from the thermal images at various heating currents $I_\mathrm{DC} =0$, 6, 10, 12, and 13~mA. The solid curves are fits used to extract the temperature gradient.
\textbf{e} In-plane temperature gradient along the $x$ axis, $\mathrm{d}T/\mathrm{d}x$, as a function of position, obtained from the temperature profile in \textbf{d}.
\textbf{f} In-plane temperature gradient $\langle \mathrm{d}T/\mathrm{d}x \rangle$, spatially averaged over the Ti/Ni junction region, as a function of $I_\mathrm{DC}^2$ for the two junctions located at $x<0$ (red) and $x>0$ (blue).
}
\label{figTiNiDevice}
\end{figure}

To observe the orbital Nernst effect, we chose Ti because its intrinsic spin Nernst effect is predicted to be negligible~\cite{PhysRevMaterials.6.095001}, which is also supported by our first-principles calculations (see Supplementary Note 1), thereby minimizing contributions from spin currents and allowing us to selectively probe signals arising from orbital currents generated by the orbital Nernst effect.
We fabricated a device comprising a Ti cross bar with Ni detection electrodes (see Fig.~\ref{figTiNiDevice}a). The Ti and Ni films, with thicknesses of 10 nm and 8 nm, respectively, were deposited by magnetron sputtering (see Methods). For the measurements of the thermoelectric response, a DC charge current $I_\mathrm{DC}$ was applied along the Ti channel ($y$ axis) to generate a lateral temperature gradient along the $x$ axis via Joule heating. In this device geometry, heat flows symmetrically from the central region of the cross bar toward the two Ni detectors, producing in-plane temperature gradients of opposite sign at the two Ti/Ni junctions (see Fig.~\ref{figTiNiDevice}b).

The in-plane temperature profile was measured by high-resolution thermography (see Methods). Because the metallic layers are thin and thermally coupled to the substrate, the measured surface profile reflects the in-plane temperature distribution in the device. The measured temperature image $T$ and the corresponding $\mathrm{d}T/\mathrm{d}x$ map directly show that the central Ti wire is heated by the applied current and that the in-plane temperature gradient has opposite signs on the two sides of the Ti wire (Fig.~\ref{figTiNiDevice}c). Figure~\ref{figTiNiDevice}d shows the temperature profiles along the $x$ axis at several heating currents, together with the fitting curves used to extract the temperature gradient. From these fits, we evaluated the position-dependent in-plane temperature gradient $\mathrm{d}T/\mathrm{d}x$ (Fig.~\ref{figTiNiDevice}e) and its value spatially averaged over each Ti/Ni junction region, $\langle \mathrm{d}T/\mathrm{d}x \rangle$, which increases linearly with $I_\mathrm{DC}^2$ as expected for Joule heating (Fig.~\ref{figTiNiDevice}f).

The device shown in Fig.~\ref{figTiNiDevice}a is designed to probe orbital currents generated by the in-plane temperature gradients induced in the Ti layers beneath the Ni electrodes. These orbital currents are detected through electrochemical potential measurements using the Ni electrodes, which act as orbital-sensitive probes due to their ability for orbital-to-spin conversion. In the device, the temperature gradient along the $x$ axis in the Ti layer generates orbital currents polarized along the $y$ axis and flowing along the $z$ axis via the orbital Nernst effect (see Fig.~\ref{figTiNiDevice}b). These orbital currents are injected into the adjacent Ni electrodes, where orbital-to-spin conversion mediated by spin-orbit coupling gives rise to spin accumulation in the Ni layer. 
Consequently, the electrochemical potential in the Ni layer away from the interface depends on the relative orientation, parallel or antiparallel, between the Ni magnetization and the polarization of the spin accumulation, namely the polarization of the injected orbital current.
For the two Ni electrodes, opposite signs of the temperature gradient lead to opposite orbital polarizations of the injected currents, while the magnetization direction remains identical (Fig.~\ref{figTiNiDevice}b).
This leads to different electrochemical potentials in the two Ni electrodes, giving rise to a measurable potential difference between them. 
This signal is maximized when the magnetization is aligned with the polarization direction of the orbital currents generated by the orbital Nernst effect, i.e., the $y$ axis, and is expected to vanish when the magnetization is perpendicular to this axis.
To detect this signal experimentally, we measured the electric potential difference $V$ between the Ni electrodes at room temperature under an in-plane magnetic field $H$ (see Methods).

\begin{figure}[tb]
\includegraphics[scale=1]{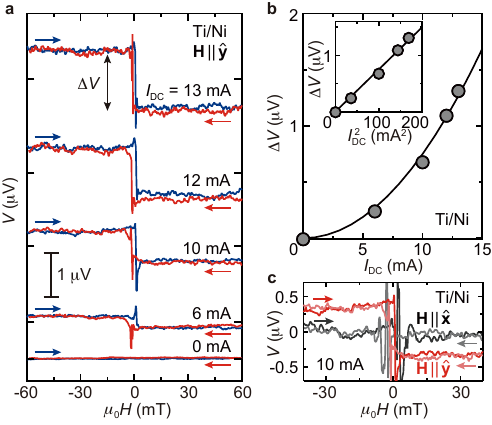}
\caption{
\textbf{Orbital Nernst effect in Ti.}
\textbf{a} Electric potential difference $V$ between the Ni electrodes in the Ti(10~nm)/Ni(8~nm) device as a function of the in-plane magnetic field $\mu_{0}H$ applied along the $y$ axis for different heating currents $I_\mathrm{DC}$, where the numbers in parentheses indicate the layer thicknesses, and $\mu_{0}$ denotes the vacuum permeability. The red and blue curves correspond to magnetic-field sweeps from positive to negative and negative to positive fields, respectively. For each dataset, a constant background voltage has been subtracted.
\textbf{b} Voltage step $\Delta V$ as a function of the heating current $I_\mathrm{DC}$ in the Ti(10~nm)/Ni(8~nm) device. The solid circles represent experimental data, and the solid curve is a quadratic fit. The inset shows $\Delta V$ plotted as a function of $I_\mathrm{DC}^2$, confirming the proportionality between $\Delta V$ and $I_\mathrm{DC}^2$.
\textbf{c} $\mu_0 H$ dependence of $V$ in the Ti(10~nm)/Ni(8~nm) device measured at $I_\mathrm{DC}=10$ mA for magnetic fields applied along the $y$ axis (${\bf H}\parallel \hat{\bf y}$, red) and the $x$ axis (${\bf H}\parallel \hat{\bf x}$, gray).
}
\label{figTiONE}
\end{figure}

\begin{figure}[tb]
\includegraphics[scale=1]{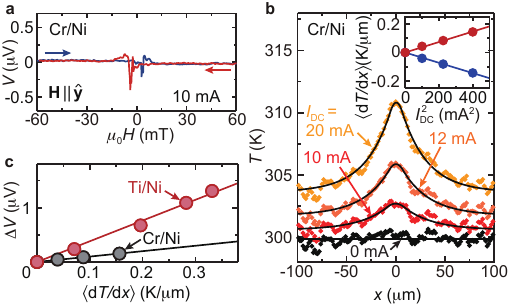}
\caption{
\textbf{Thermoelectric response in Cr/Ni.}
\textbf{a} $\mu_0 H$ dependence of $V$ in the Cr(10~nm)/Ni(8~nm) device measured at $I_\mathrm{DC}=10$~mA, with the magnetic field applied along the $y$ axis ($\mathbf{H}\parallel \hat{\mathbf{y}}$).
\textbf{b} Temperature profiles $T$ along the $x$ axis measured by thermography in the Cr(10~nm)/Ni(8~nm) device at $I_\mathrm{DC}=0$, 10, 12, and 20~mA. The inset shows the in-plane temperature gradient $\langle \mathrm{d}T/\mathrm{d}x \rangle$ spatially averaged over the Cr/Ni junction region as a function of $I_\mathrm{DC}^2$ for the two junctions located at $x<0$ (red) and $x>0$ (blue).
\textbf{c} Voltage step $\Delta V$ as a function of the in-plane temperature gradient $\langle \mathrm{d}T/\mathrm{d}x \rangle$ for the Ti(10~nm)/Ni(8~nm) (red) and Cr(10~nm)/Ni(8~nm) (gray) devices.
Here, $\langle \mathrm{d}T/\mathrm{d}x \rangle$ denotes the in-plane temperature gradient in the Ti or Cr layer, spatially averaged over the Ti/Ni or Cr/Ni junction region and measured by thermography at each $I_\mathrm{DC}$.
}
\label{figCrNi}
\end{figure}

Figure~\ref{figTiONE}a shows $V$ as a function of $H$ applied along the $y$ axis (${\bf H}\parallel \hat{\bf y}$) for different heating currents $I_\mathrm{DC}$, where a constant background voltage has been subtracted. The results reveal clear voltage steps around $\mu_0 H=0$~mT, indicating that a magnetization-dependent electric potential difference is induced between the Ni electrodes. The voltage step $\Delta V$, defined as the difference between the saturated voltage levels at positive and negative fields (see Methods), increases quadratically with the heating current $I_\mathrm{DC}$, as shown in Fig.~\ref{figTiONE}b, evidencing its thermal origin. 
Notably, as shown in Fig.~\ref{figTiONE}c, $\Delta V$ vanishes when the magnetic field is applied along the $x$ axis (${\bf H}\parallel \hat{\bf x}$), consistent with the expectation that it arises from orbital currents polarized along the $y$ axis.
These observations indicate that the temperature gradient along the $x$ axis induces spin accumulation polarized along the $y$ axis in the Ni layers, in agreement with the prediction of the orbital Nernst effect in the Ti layer combined with orbital-to-spin conversion in the Ni electrodes.

\vspace{12pt}
\noindent
\textbf{Thermoelectric contributions from Ni}

The thermally induced voltage $\Delta V$ could in principle contain a contribution from the anomalous Nernst effect in the Ni electrodes. To test this possibility, we measured the heat-induced voltage in a Cr/Ni device, in which the Ti layer was replaced with Cr. Figure~\ref{figCrNi}a shows the magnetic-field $H$ dependence of $V$ between the Ni electrodes under a heating current $I_\mathrm{DC}$ in the Cr/Ni device, measured in the same orbital Nernst configuration used for the Ti/Ni device, namely the voltage between the two Ni electrodes with $H$ applied along the $y$ axis (Figs.~\ref{figTiNiDevice}a and \ref{figTiNiDevice}b). In sharp contrast to the Ti/Ni device, the Cr/Ni device exhibits negligible $\Delta V$. We determined the in-plane temperature gradient in the Cr/Ni device from thermography measurements, following the same procedure as that used for the Ti/Ni device (Fig.~\ref{figCrNi}b). 

As shown in Fig.~\ref{figCrNi}c, $\Delta V$ in the Ti/Ni device increases linearly with the in-plane temperature gradient $\langle \mathrm{d}T/\mathrm{d}x \rangle$, whereas it is strongly suppressed in the Cr/Ni device. Because the Ni detectors and measurement geometry are nominally identical in the Ti/Ni and Cr/Ni devices, any contribution intrinsic to Ni would be expected to produce comparable responses in the two devices. The pronounced difference therefore demonstrates that the signal observed in the Ti/Ni device is not dominated by thermoelectric effects generated within the Ni detector, including anomalous Nernst voltages or possible self-induced signals in Ni~\cite{LiuZhu2025}.

\begin{figure}[tb]
\includegraphics[scale=1]{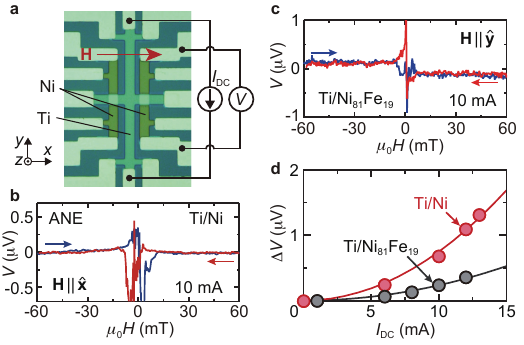}
\caption{
\textbf{Anomalous Nernst contribution and ferromagnetic-electrode dependence of orbital Nernst signal.}
\textbf{a} Optical image of the Ti/Ni device with the measurement configuration for detecting the anomalous Nernst effect induced by an out-of-plane temperature gradient. The red arrow indicates the direction of the applied magnetic field $\mathbf{H}$.
\textbf{b} $\mu_0 H$ dependence of $V$ between the ends of the Ni wire (see \textbf{a}) in the Ti(10 nm)/Ni(8 nm) device at $I_\mathrm{DC} = 10$ mA for ${\bf H} \parallel \hat{\mathbf{x}}$.
\textbf{c} Electric potential difference $V$ between the Ni$_{81}$Fe$_{19}$ electrodes in the Ti(10~nm)/Ni$_{81}$Fe$_{19}$(8~nm) device as a function of in-plane magnetic field $\mu_0 H$ applied along the $y$ axis under $I_\mathrm{DC}=10$ mA. The measurements were performed in the configuration shown in Fig.~\ref{figTiNiDevice}a.
\textbf{d} Voltage step $\Delta V$ as a function of heating current $I_\mathrm{DC}$ for the Ti(10~nm)/Ni(8~nm) (red) and Ti(10~nm)/Ni$_{81}$Fe$_{19}$(8~nm) (gray) devices.
}
\label{figTiPy}
\end{figure}

To further quantify any residual contribution from the anomalous Nernst effect, we performed direct control measurements in the Ti/Ni device. These measurements were designed to detect spurious anomalous Nernst voltages that could arise from unintended temperature gradients in the Ni electrodes induced by the heating current $I_\mathrm{DC}$. In the anomalous Nernst effect, the induced electric field is perpendicular to both the magnetization and the temperature gradient~\cite{Adachi2025}. We denote anomalous-Nernst voltages as $V_\mathrm{ANE}^{i,\nabla T_j}$, where $i$ indicates the voltage direction and $j$ indicates the direction of the temperature gradient driving the anomalous Nernst effect. To evaluate the contribution from an out-of-plane temperature gradient, $V_\mathrm{ANE}^{y,\nabla T_z}$, we applied $H$ along the $x$ direction while driving $I_\mathrm{DC}$ and measured the voltage along the $y$ axis between the ends of the Ni wire (Fig.~\ref{figTiPy}a). In this configuration, an out-of-plane temperature gradient in the Ni layer would generate a voltage along the $y$ axis via the anomalous Nernst effect. However, as shown in Fig.~\ref{figTiPy}b, the measured voltages for positive and negative fields are nearly identical, indicating that the anomalous Nernst contribution from the out-of-plane temperature gradient is negligible.

We also evaluated the anomalous Nernst voltage $V_\mathrm{ANE}^{z,\nabla T_{x}}$ that could be generated by an in-plane temperature gradient. In the orbital Nernst measurement geometry, an in-plane temperature gradient along $x$ in the Ni layer and magnetization along $y$ can generate an out-of-plane voltage along $z$, which can affect the electrochemical potential at the top surface of the Ni electrode. To quantify this contribution, we measured $V_\mathrm{ANE}^{y,\nabla T_{x}}$ by applying an out-of-plane magnetic field along $z$ and detecting the voltage along the $y$ axis between the ends of the Ni wire while driving $I_\mathrm{DC}$. We converted the measured voltage into the out-of-plane voltage expected in the orbital Nernst measurement geometry using the geometry of the Ni wire (see Supplementary Note 2). At $I_\mathrm{DC}=10$ mA, the estimated value of $V_\mathrm{ANE}^{z,\nabla T_{x}}$ is approximately 60 nV for the Ti/Ni device, which is negligible compared with the measured signal $\Delta V \simeq 680$~nV shown in Fig.~\ref{figTiONE}a. These results confirm that parasitic anomalous Nernst voltages in the Ni layer do not account for the observed $\Delta V$ in the Ti/Ni device, indicating that the observed voltage is driven by an angular-momentum current generated in the Ti layer rather than by a thermoelectric response within the Ni detector.

\vspace{12pt}
\noindent
\textbf{Distinguishing orbital and spin Nernst contributions}

Another possible contribution to the observed $\Delta V$ is the spin Nernst effect in Ti. Although its intrinsic spin Nernst conductivity is predicted to be negligible (see Supplementary Note 1), we test this possibility experimentally by examining the detector-material dependence of $\Delta V$. 
Specifically, we measured $V$ as a function of $H$ applied along the $y$ axis in a device in which Ni was replaced with Ni$_{81}$Fe$_{19}$, as shown in Fig.~\ref{figTiPy}c, using the same orbital Nernst measurement configuration as for the Ti/Ni device (Fig.~\ref{figTiNiDevice}a).
To establish the detector dependence expected from the spin Nernst effect, we consider the drift-diffusion model of spin injection and detection in a nonmagnetic/ferromagnetic structure, in which the voltage step is expressed as~\cite{bose2018direct}
\begin{widetext}
\begin{equation}
\Delta V= \frac{2P j_\mathrm{S} \lambda_\mathrm{FM} \lambda_\mathrm{NM}\left[\cosh ({t_\mathrm{NM}}/{\lambda_\mathrm{NM}})-1\right]}{\lambda_\mathrm{NM} (1-P^2)\sigma_\mathrm{FM} \cosh ({t_\mathrm{NM}}/{\lambda_\mathrm{NM}})+\lambda_\mathrm{FM} \sigma_\mathrm{NM} \sinh ({t_\mathrm{NM}}/{\lambda_\mathrm{NM}})}.\label{eq:SNE}
\end{equation}
\end{widetext}
Here, $P=(\sigma_\uparrow - \sigma_\downarrow)/(\sigma_\uparrow + \sigma_\downarrow)$ and $j_\mathrm{S}$ are the spin polarization of the ferromagnetic layer and the spin current density generated in the nonmagnetic layer, respectively. $\sigma_{\uparrow(\downarrow)}$ is the conductivity of the majority (minority) spin channel in the ferromagnetic layer. $\lambda_\mathrm{FM(NM)}$ and $\sigma_\mathrm{FM(NM)}$ are the spin diffusion length and conductivity of the ferromagnetic (nonmagnetic) layers, respectively, and $t_\mathrm{NM}$ is the thickness of the nonmagnetic layer. Equation~(\ref{eq:SNE}) shows that $\Delta V$ is governed mainly by $P$, $\lambda_\mathrm{FM}$, and $\sigma_\mathrm{FM}$. The reported spin-diffusion lengths of Ni and Ni$_{81}$Fe$_{19}$ are comparable~\cite{dx73-2lp6,ko2020optical}. In contrast, Ni$_{81}$Fe$_{19}$ has a larger spin polarization than Ni~\cite{PhysRevB.67.085319} and a lower conductivity, both of which increase $\Delta V$. Thus, $\Delta V$ arising from the spin Nernst effect is expected to be larger in the Ti/Ni$_{81}$Fe$_{19}$ device than in the Ti/Ni device. Experimentally, however, $\Delta V$ is strongly suppressed rather than enhanced when the Ni detector is replaced with Ni$_{81}$Fe$_{19}$, as shown in Fig.~\ref{figTiPy}d. This detector dependence therefore indicates that the spin Nernst effect is not the dominant origin of the observed signal.

By contrast, this detector dependence provides key experimental evidence for an orbital Nernst origin. In an orbital Nernst scenario, the ferromagnetic detection electrodes play a crucial role by converting the orbital currents generated by the orbital Nernst effect in Ti into spin accumulation, which is detected electrically as a voltage. 
Notably, a similar ferromagnet-based detection principle was recently established in nonlocal orbital Edelstein measurements, where the strong suppression of the signal with Ni$_{81}$Fe$_{19}$ detectors provided key evidence for an orbital origin~\cite{GaoOEE2025}.
When an orbital current is injected into a ferromagnet, the orbital angular momentum can be transferred to the lattice through crystal-field torque and to the spin system through spin-orbit coupling (see Supplementary Note 1). The resulting spin accumulation reflects the competition between orbital quenching and orbital-to-spin transfer. Rather than being determined solely by the atomic spin-orbit coupling strength, orbital-to-spin conversion is governed by the spin-orbit correlation near the Fermi energy, which is sensitive to the electronic structure and ferromagnet composition~\cite{PhysRevResearch.2.013177}. Among conventional 3$d$ ferromagnets, Ni is predicted to exhibit the strongest orbital-to-spin conversion~\cite{Ta-orbital}. Fe doping in Ni, however, has been theoretically shown to suppress the spin-orbit correlation near the Fermi energy~\cite{Lee2024CAP}, providing a microscopic basis for weaker orbital responses in Ni$_{81}$Fe$_{19}$ than in Ni reported previously~\cite{hayashi2022observation,PhysRevResearch.5.023054,moriya2024nano,10.1063/5.0263240,PhysRevB.106.184406,seifert2023time,GaoOEE2025}. 
Thus, the strong suppression of $\Delta V$ in the Ti/Ni$_{81}$Fe$_{19}$ device is consistent with the detector-material dependence expected for an orbital response, while being opposite to the trend expected for a spin Nernst response, supporting the conclusion that the observed voltage is dominated by the orbital Nernst effect.

The strong suppression of $\Delta V$ in the Cr/Ni device (Fig.~\ref{figCrNi}c), while primarily ruling out a dominant thermoelectric contribution from the Ni detector, also provides a complementary test of a spin Nernst interpretation. 
The calculated intrinsic thermal spin Nernst conductivity at about 300~K is much larger in magnitude in Cr than in Ti: $\alpha_\mathrm{SNE}^\mathrm{T}=-0.0605\ (\hbar/e)\ \mathrm{A}\,\mathrm{K}^{-1}\,\mathrm{m}^{-1}$ for Cr and $0.0165\ (\hbar/e)\ \mathrm{A}\,\mathrm{K}^{-1}\,\mathrm{m}^{-1}$ for Ti (see Supplementary Note 1). Because both the Seebeck coefficient and the magnitude of the spin Hall conductivity are greater in Cr than in Ti, the thermoelectric spin Nernst contribution is also expected to be stronger in Cr. A dominant intrinsic spin Nernst contribution would therefore predict a stronger response in the Cr/Ni device than in the Ti/Ni device. The experimentally observed strong suppression in the Cr/Ni device is contrary to this prediction, providing additional support for the conclusion that the spin Nernst effect is not the dominant origin of the signal in the Ti/Ni device.

The comparison between the Ti/Ni and Cr/Ni devices, the anomalous-Nernst controls, and the ferromagnetic-detector dependence therefore consistently indicate that the heat-induced signal $\Delta V$ in the Ti/Ni device is dominated by the orbital Nernst effect in Ti followed by orbital-to-spin conversion in Ni, rather than by the spin Nernst effect or thermoelectric effects in the Ni layer. We note that this conclusion is based primarily on the experimental trends, with the calculated intrinsic spin Nernst conductivities providing complementary support.
The orbital-Nernst-effect interpretation is further supported by the Ti-thickness dependence of the signal: the heating-power-normalized voltage step tends to increase with increasing Ti thickness (see Supplementary Note 3). This thickness dependence supports a bulk origin in the Ti layer, rather than a purely interfacial effect, consistent with the orbital Nernst effect.

\vspace{12pt}
\noindent 
\textbf{Dominant contribution to the orbital Nernst response}

With the orbital Nernst effect identified as the dominant origin of the observed signal, we next examine the relative importance of the thermal and thermoelectric contributions to the orbital Nernst response in Ti.
Analogous to the spin Nernst effect~\cite{MA2010510,PhysRevLett.109.026601}, the orbital Nernst conductivity, $\alpha_\mathrm{ONE}$, may be decomposed into thermal and thermoelectric contributions: $\alpha_\mathrm{ONE} = \alpha_\mathrm{ONE}^\mathrm{T} + \alpha_\mathrm{ONE}^\mathrm{TE}$. The thermal contribution, $\alpha_\mathrm{ONE}^\mathrm{T}$, arises purely from the temperature gradient and is governed by the energy derivative of the orbital Hall conductivity $\sigma_\mathrm{OHE}$ according to the Mott relation~\cite{PhysRevMaterials.6.095001}:
\begin{equation}
\alpha_\mathrm{ONE}^\mathrm{T} = -\frac{\pi^2 k_\mathrm{B}^2 T}{3 e}
\left(\frac{\partial \sigma_\mathrm{OHE}}{\partial E}\right)_{E=E_\mathrm{F}}, \label{eq:ONE}
\end{equation}
where $k_\mathrm{B}$, $e$, and $E_\mathrm{F}$ are the Boltzmann constant, the elementary charge, and the Fermi energy, respectively. This contribution reflects the variation of the orbital Hall conductivity near the Fermi energy and is independent of the Seebeck coefficient $S$. In contrast, the thermoelectric contribution, $\alpha_\mathrm{ONE}^\mathrm{TE}$, arises from the interplay between the Seebeck effect and the orbital Hall effect~\cite{MA2010510}: $\alpha_\mathrm{ONE}^\mathrm{TE} = S\sigma_\mathrm{OHE}$.
Accordingly, the voltage step due to the orbital Nernst effect can be expressed as $\Delta V_\mathrm{ONE} = j_\mathrm{L}^\mathrm{ONE}  \xi  =  \alpha_\mathrm{ONE}(-\langle \mathrm{d}T/\mathrm{d}x \rangle)  \xi=-(\alpha_\mathrm{ONE}^\mathrm{T} + \alpha_\mathrm{ONE}^\mathrm{TE}) \langle \mathrm{d}T/\mathrm{d}x \rangle  \xi$, where $j_\mathrm{L}^\mathrm{ONE}$ is the orbital current generated by the orbital Nernst effect, $\langle \mathrm{d}T/\mathrm{d}x \rangle$ is the in-plane temperature gradient in the Ti layer spatially averaged over the Ti/Ni junction region, and $\xi$ is a phenomenological parameter that characterizes the conversion efficiency from $j_\mathrm{L}^\mathrm{ONE}$ into $\Delta V_\mathrm{ONE}$. 
At this stage, quantifying $\xi$ is challenging owing to the limited quantitative understanding of orbital transport across the interface and orbital-to-spin conversion within the ferromagnetic layer, which precludes a direct estimation of $\alpha_\mathrm{ONE}$ from the measured $\Delta V$.
Nevertheless, the voltage step due to the thermoelectric contribution, $\Delta V_\mathrm{ONE}^\mathrm{TE} = -\alpha_\mathrm{ONE}^\mathrm{TE} \langle \mathrm{d}T/\mathrm{d}x \rangle  \xi = -S \sigma_\mathrm{OHE} \langle \mathrm{d}T/\mathrm{d}x \rangle \xi$, can be estimated by independently measuring the voltage step induced by the orbital Hall effect, as described below.

\begin{figure}[tb]
\includegraphics[scale=1]{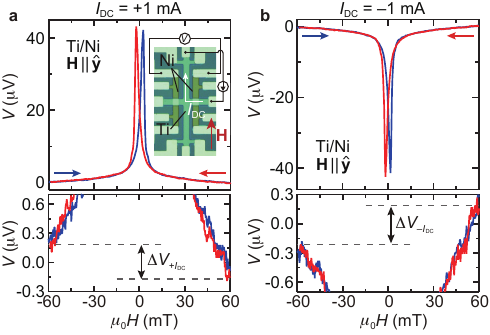}
\caption{
\textbf{Orbital Hall effect in Ti.}
Electric potential difference $V$ between the Ni electrodes in the Ti(10~nm)/Ni(8~nm) device as a function of the in-plane magnetic field $\mu_0 H$ applied along the $y$ axis under ({\bf a}) a positive current $I_\mathrm{DC}=+1$ mA and ({\bf b}) a negative current $I_\mathrm{DC}=-1$ mA. For each dataset, a constant background voltage has been subtracted. The red and blue curves correspond to opposite field-sweep directions. 
The inset to ({\bf a}) shows an optical microscope image of the Ti/Ni device together with the measurement configuration for the orbital Hall effect. An in-plane charge current $I_\mathrm{DC}$ is applied along the Ti layer, and the voltage $V$ is measured across the Ni electrodes.
The lower panels provide magnified views, highlighting the voltage differences at positive and negative fields, defined as $\Delta V_{+I_\mathrm{DC}}$ for $+I_\mathrm{DC}$ and $\Delta V_{-I_\mathrm{DC}}$ for $-I_\mathrm{DC}$.
}
\label{figOHE}
\end{figure}

To identify the primary contribution of the orbital Nernst effect to the observed $\Delta V$ in the Ti/Ni device, we measured the orbital Hall effect using the same device. For this measurement, a DC in-plane charge current $I_\mathrm{DC}$ was applied along the Ti/Ni layer (see the inset to Fig.~\ref{figOHE}a). In this configuration, the charge current drives the orbital Hall effect in the Ti layer, generating a transverse orbital current $j_\mathrm{L}^\mathrm{OHE} = \sigma_\mathrm{OHE} E$, where $E$ denotes the electric field. The generated orbital current is injected into the adjacent Ni layer and converted into a measurable spin accumulation through orbital-to-spin conversion, resulting in a voltage step $\Delta V _\mathrm{OHE}= j_\mathrm{L}^\mathrm{OHE}  \xi = \sigma_\mathrm{OHE} E  \xi$, under the assumption that the coefficient $\xi$ is identical for the orbital Nernst and orbital Hall processes.

Figures~\ref{figOHE}a and \ref{figOHE}b show the voltage $V$ signals measured between the two Ni electrodes in the Ti/Ni device after subtraction of a constant background. The results exhibit a clear voltage difference between positive and negative magnetic fields, $\Delta V_{+I_\mathrm{DC}} \simeq 500$~nV at $I_\mathrm{DC}=1$~mA, where the applied current is sufficiently small to neglect any orbital Nernst contribution arising from Joule heating (see also Fig.~\ref{figTiONE}b).
When the charge current direction is reversed, the sign of this voltage difference also reverses (see the lower panels in Figs.~\ref{figOHE}a and \ref{figOHE}b), consistent with the prediction of the orbital Hall effect. In this geometry, the in-plane charge current in the Ni layer with in-plane magnetization can generate a potential difference between the top and bottom Ni surfaces via the anomalous Hall effect, providing an additional contribution to the measured voltage. This contribution is estimated to be $\Delta V_{+I_\mathrm{DC}}^\mathrm{AHE} \simeq 180$~nV at $I_\mathrm{DC}=1$~mA (see Methods), yielding an orbital Hall contribution of $\Delta V_{+I_\mathrm{DC}}^\mathrm{OHE} \simeq 320$~nV.

A comparison between the orbital Nernst and orbital Hall signals allows us to clarify the dominant contribution to the observed orbital Nernst effect. 
In the orbital Nernst geometry, orbital currents are injected into both Ni electrodes, whereas in the orbital Hall geometry the current flows into only one electrode, leading to $\Delta V_\mathrm{OHE} = 2\Delta V_{+I_\mathrm{DC}}^\mathrm{OHE} = \sigma_\mathrm{OHE} E \xi$. Using the measured values of $\Delta V_{+I_\mathrm{DC}}^\mathrm{OHE}$ and $E$ together with the Seebeck coefficient~\cite{alasli2021high} $S=7.6$~$\mu$V/K, we estimate the thermoelectric contribution to the orbital Nernst signal as $\Delta V_\mathrm{ONE}^\mathrm{TE} =- S \sigma_\mathrm{OHE} \langle \mathrm{d}T/\mathrm{d}x \rangle \xi = -2S \langle \mathrm{d}T/\mathrm{d}x \rangle \Delta V_{+I_\mathrm{DC}}^\mathrm{OHE}/E \simeq  -0.4$~nV, where the in-plane temperature gradient spatially averaged over the Ti/Ni junction region of the device $\langle \mathrm{d}T/\mathrm{d}x \rangle$ = 0.19~K/$\mu$m at $I_\mathrm{DC} = 10$~mA was determined from the thermography measurement (see Fig.~\ref{figTiNiDevice}f). This estimated thermoelectric contribution is orders of magnitude smaller than the experimentally observed voltage step, $\Delta V \simeq 680$~nV at $I_\mathrm{DC} = 10$~mA (Fig.~\ref{figTiONE}b), indicating that the thermoelectric contribution $\alpha_\mathrm{ONE}^\mathrm{TE}$ plays only a minor role in the observed signal. These results show that the thermal contribution $\alpha_\mathrm{ONE}^\mathrm{T}$, purely driven by the temperature gradient, dominates the observed orbital Nernst effect. 

The comparison between the Ti/Ni and Cr/Ni devices provides additional support for the conclusion that the thermal contribution $\alpha_\mathrm{ONE}^\mathrm{T}$ dominates the observed orbital Nernst effect. The orbital Hall conductivity of Cr at $E_\mathrm{F}$ has been predicted to be comparable to that of Ti~\cite{PhysRevMaterials.6.095001}. Using the Seebeck coefficient of Cr~\cite{rowe1995crc}, $S=21.8~\mu\mathrm{V/K}$, we find that the thermoelectric contribution $\alpha_\mathrm{ONE}^\mathrm{TE}$, associated with the Seebeck effect and the orbital Hall effect, is negligible, as in the Ti/Ni device. These results suggest that the key difference between Ti and Cr in generating $\Delta V$ lies in the orbital Nernst response driven directly by the temperature gradient, $\alpha_\mathrm{ONE}^\mathrm{T}$. In Cr, the orbital Hall conductivity is nearly insensitive to energy near $E_\mathrm{F}$, whereas it varies significantly in Ti. Accordingly, the calculated intrinsic orbital Nernst conductivity of Cr at about 300~K, $\alpha_\mathrm{ONE}^\mathrm{T}=-0.0391\ (\hbar/e)\ \mathrm{A}\,\mathrm{K}^{-1}\,\mathrm{m}^{-1}$, is much smaller than that of Ti, $\alpha_\mathrm{ONE}^\mathrm{T}=-0.730\ (\hbar/e)\ \mathrm{A}\,\mathrm{K}^{-1}\,\mathrm{m}^{-1}$, reflecting the smaller $\partial \sigma_\mathrm{OHE}/\partial E$ through the Mott relation (see Supplementary Note 1). The strong suppression of $\Delta V$ in the Cr/Ni device is therefore consistent with the calculated trend of the orbital Nernst conductivity.
We note that possible extrinsic contributions, such as extrinsic spin and orbital Nernst effects and interface scattering, cannot be fully assessed in the present study. A more quantitative analysis incorporating these contributions will be the subject of future work.

\vspace{12pt}
\noindent 
\textbf{Conclusions and outlook}

We have experimentally observed thermally induced voltage signals consistent with the orbital Nernst effect, the thermal analogue of the orbital Hall effect, thereby establishing a previously missing element in the framework of spin and orbital caloritronic phenomena. Systematic measurements of the magnetic-field and ferromagnetic-detector dependences, together with the comparison between the Ti/Ni and Cr/Ni devices and controls for anomalous Nernst contributions, identify the orbital Nernst effect as the dominant origin of the observed response.
These results provide a robust experimental foundation for exploring thermally driven orbital transport and are anticipated to stimulate further experimental and theoretical studies, opening new avenues for orbital caloritronics, which bridges orbitronics, spintronics, and thermoelectrics.

\clearpage

\bigskip\noindent
\textbf{Methods}\\
\textbf{Device fabrication.}
To investigate the orbital Nernst effect in Ti, we fabricated devices consisting of a Ti cross bar and ferromagnetic Ni electrodes on a thermally oxidized Si substrate with a 100-nm-thick SiO$_2$ layer. The Si substrate and the device layers were electrically isolated from each other by the SiO$_2$ layer. All films were deposited by magnetron sputtering at a base pressure better than $1\times10^{-6}$ Pa. The 10-nm-thick Ti cross bar was patterned by photolithography and lift-off, and capped with a 1-nm-thick Pt layer to prevent oxidation. Subsequently, 8-nm-thick Ni wires were defined by photolithography and lift-off. Prior to Ni deposition, the Pt capping layer was removed by Ar ion milling to ensure direct Ti/Ni contact. Finally, Ti(5 nm)/Cu(300 nm) electrodes were deposited to form electrical contacts, where the numbers in parentheses indicate thickness. The Ti cross bar and Ni wires each had a width of 10~$\mu$m, giving a Ti/Ni junction cross section of $10~\mu\mathrm{m}\times10~\mu\mathrm{m}$. The separation between the Ti and Ni wires was 10~$\mu$m.
The Ti/Ni$_{81}$Fe$_{19}$ and Cr/Ni devices were fabricated by the same process as the Ti/Ni devices.
The crystal structure of the Ti layer was examined by transmission electron microscopy (TEM) and X-ray diffraction (XRD) (see Supplementary Note 4).

\noindent
\\\textbf{Voltage measurements.}
We measured the electric potential difference $V$ between the Ni electrodes as a function of an in-plane magnetic field $H$ while applying a current $I_\mathrm{DC}$ at room temperature. In the orbital Nernst measurements, the heating current was applied to the Ti cross bar along the $y$ axis, and the in-plane magnetic field was applied along either the $x$ or $y$ axis. To extract thermally induced signals, the measured voltage was averaged between measurements at positive ($+I_\mathrm{DC}$) and negative ($-I_\mathrm{DC}$) heating currents. The data were further averaged over multiple measurements to improve the signal-to-noise ratio. 
A pressure of $5\times10^{-3}$ Pa was maintained in the sample space during the experiments.
The voltage step $\Delta V$ was defined as the difference between the average voltages measured in the field ranges $-60$ to $-20$ mT and 20 to 60 mT.

In principle, the two Ti/Ni junctions generate orbital-related changes in electrochemical potential with opposite signs, because the in-plane temperature gradients at the two junctions have opposite signs (Fig.~\ref{figTiNiDevice}b). We examined single-junction voltage measurements between one Ni detector and the Ti electrode. In this geometry, however, the heating current produces a large background voltage through the Ti channel, of the order of several volts, which is approximately six orders of magnitude larger than the orbital-Nernst-related voltage detected in the two-contact configuration. The field-dependent component associated with a single junction therefore cannot be extracted reliably. We consequently use the differential voltage between the two Ni electrodes, which excludes the large heating-current voltage drop and directly detects the difference between the two junction responses.

In the orbital Hall measurements, the in-plane charge current in the Ni layer with in-plane magnetization can generate an additional voltage via the anomalous Hall effect. To evaluate this contribution, we measured the in-plane voltage $\Delta V_{\mathrm{AHE}}^{\mathrm{in}}$ in a Ti/Ni cross structure with a length and a width of $l=w=10~\mu\mathrm{m}$ while applying $I_\mathrm{DC}=1$~mA under an out-of-plane magnetic field along the $z$ direction, which fully saturates the magnetization in the device. Taking the geometrical factor into account, the out-of-plane voltage due to the anomalous Hall effect under the orbital Hall measurement conditions is obtained as $\Delta V^{\mathrm{AHE}}_{+I_\mathrm{DC}}=\rho_{\mathrm{Ni}}^{\mathrm{out}}\Delta V_{\mathrm{AHE}}^{\mathrm{in}}/(R^{\mathrm{in}}_{\mathrm{Ti/Ni}}l)=180$~nV, where $\rho_{\mathrm{Ni}}^{\mathrm{out}}=7~\mu\Omega\mathrm{cm}$ is the out-of-plane resistivity of the Ni layer, estimated from the thickness dependence of the resistivity, and $R^{\mathrm{in}}_{\mathrm{Ti/Ni}}=19~\Omega$ is the in-plane resistance of the Ti/Ni cross. This value is subtracted from the total measured signal to extract the net orbital Hall contribution.

\noindent
\\\textbf{Measurement of temperature profile.}
The temperature distribution in the devices was measured by high-resolution infrared thermography (ELITE System). Here, to enhance the infrared emissivity and ensure uniform emission properties, the top surface of the devices was coated with insulating black ink with high infrared emissivity (JSC-3, Japansensor Corporation). The relation between the infrared intensity and temperature was calibrated for this black ink. The thermal images were acquired over the viewing area of $176 \times 400$ pixels with a pixel size of 1.5~$\mu$m. 
During the thermography measurements, a heating current $I_\mathrm{DC}$ was applied along the $y$ direction to the central Ti or Cr line (Fig.~\ref{figTiNiDevice}a).
To extract the position-dependent in-plane temperature gradient $\mathrm{d}T/\mathrm{d}x$ and its value $\langle \mathrm{d}T/\mathrm{d}x \rangle$ spatially averaged over the Ti/Ni or Cr/Ni junction region, line profiles along the $x$ axis were fitted using a one-dimensional heat diffusion model with a finite-width heat source:
\begin{equation}
T(x)=T_0+a \int_{-w / 2}^{w / 2} \exp \left(-\frac{|x-\xi|}{\lambda}\right) d \xi,
\end{equation}
where $T_0$, $a$, $w$, and $\lambda$ are fitting parameters, with $w$ and $\lambda$ denoting the effective width of the heat source and the characteristic heat diffusion length, respectively.

The thermal conductivity of the Si substrate, which most significantly affects the heat flow, was measured using the time-domain thermoreflectance technique~\cite{https://doi.org/10.1002/adfm.202506554, 10.1063/1.1819431}, yielding a value of $138 \pm 11$ W/mK.

\noindent
\\\textbf{Structural characterization.}
The crystal structure of the Ti layer was characterized by cross-sectional TEM and XRD measurements. Scanning transmission electron microscopy (STEM) observations were carried out using a probe-aberration-corrected Titan G2 80--200 microscope operated at 200 kV. Thin lamellae for STEM analysis were prepared using a focused ion beam/scanning electron microscope (FIB-SEM, FEI Helios G5) via the lift-out technique. XRD measurements were performed using Cu K$\alpha$ radiation. The TEM and XRD analyses show that the Ti layer has an fcc structure with a (111) orientation (see Supplementary Note 4).

\vspace{20pt}

\noindent\textbf{Data availability}\\
The data that support the findings of this study are available within the paper and the Supplementary Information.
Source data are provided with this paper.

\clearpage

\textbf{References}\\

\clearpage

\bigskip\noindent
Correspondence and requests for materials should be addressed to K.A. (ando@appi.keio.ac.jp)\\

\bigskip\noindent
\textbf{Acknowledgements}\\
This work was supported by JSPS KAKENHI (Grant Numbers: 25K21707 (K.A.), 22H04964 (K.A.), and 22H04965 (K.U.)), Spintronics Research Network of Japan (Spin-RNJ) (K.A.), MEXT Initiative to Establish Next-generation Novel Integrated Circuits Centers (X-NICS) (Grant Number: JPJ011438) (K.A.), and JST ERATO ``Magnetic Thermal Management Materials" (Grant Number: JPMJER2201) (K.U.).
This work was further supported by the Knut and Alice Wallenberg Foundation (Grant Nos. 2022.0079 and 2023.0336) (P.M.O.), the Wallenberg Initiative Materials Science for Sustainability (WISE) funded by the Knut and Alice Wallenberg Foundation (D.J. and P.M.O.), and the EIC Pathfinder OPEN ``OBELIX'' (Grant No. 101129641) (P.M.O.). The calculations were supported by resources provided by the National Academic Infrastructure for Supercomputing in Sweden (NAISS) at NSC Link\"oping, partially funded by Vetenskapsr{\aa}det through Grant Nos. 2022-06725 and 2026-05211.

\bigskip\noindent
\textbf{Author contributions}\\
Y.M. fabricated the devices, performed the measurements, and analyzed the data in collaboration with N.Y.
Y.M., K.U., and K.A. developed the interpretation of the experimental results.
T.H. and K.U. performed the thermography measurements and thermal conductivity measurements.
D.J. and P.M.O. performed the first-principles calculations.
H.S.-A. performed the TEM analysis.
Y.M. and K.A. wrote the manuscript with input from N.Y., T.H., D.J., P.M.O., H.S.-A., and K.U.
K.A. supervised the project. All authors discussed the results and reviewed the manuscript.

\bigskip\noindent
\textbf{Competing interest}\\
The authors declare no competing interests.

\end{document}